# OUR PRACTICE OF USING MACHINE LEARNING TO RECOGNIZE SPECIES BY VOICE

Siddhardha Balemarthy
*Deakin University*
Melbourne, Australia
sbalemar@deakin.edu.au

Atul Sajjanhar
*Deakin University*
Melbourne, Australia
atul.sajjanhar@deakin.edu.au

Xi Zheng
*Macquarie University*
Sydney, Australia
james.zheng@mq.edu.au

*Abstract*— As the technology is advancing, audio recognition in machine learning is improved as well. Research in audio recognition has traditionally focused on speech. Living creatures (especially the small ones) are part of the whole ecosystem, monitoring as well as maintaining them are important tasks. Species such as animals and birds are tending to change their activities as well as their habitats due to the adverse effects on the environment or due to other natural or man-made calamities. For those in far deserted areas, we will not have any idea about their existence until we can continuously monitor them. Continuous monitoring will take a lot of hard work and labor. If there is no continuous monitoring, then there might be instances where endangered species may encounter dangerous situations. The best way to monitor those species are through audio recognition. Classifying sound can be a difficult task even for humans. Powerful audio signals and their processing techniques make it possible to detect audio of various species. There might be many ways wherein audio recognition can be done. We can train machines either by pre-recorded audio files or by recording them live and detecting them. The audio of species can be detected by removing all the background noise and echoes. Smallest sound is considered as a syllable. Extracting various syllables is the process we are focusing on which is known as audio recognition in terms of Machine Learning (ML).

ML algorithms are used to train the machines by considering labelled datasets of previously known species of interest. Mel-frequency cepstral coefficients are extracted from the audio datasets. Further after extracting those audio datasets, we will follow mean computation along with various k-nearest neighbor classifiers.

Audio event detection (AED) is defined as analyzing a continuous acoustic signal to extract the sound events present in the acoustic scene. The system is designed with machine learning algorithm (Hidden Markov Model) using Tensor Flow. The manuscript mainly serves as a technical paper showing how we implemented such system to achieve satisfactory result.

## I. Introduction

We need to make sure the animals and birds living in the forest are well protected for various reasons. 80% of biodiversity on land lives in the forests which signifies the importance it has in maintaining the ecological balance. It is also important to understand how the ecological footprint of human is effecting the animals and their ecosystem in the forest. This method not only would be helpful in finding how the animals are effected by our footprint but also can help us warn about a lot of natural calamities beforehand.

As maintaining a proper balance between humans and other species, we need to make sure we understand their problems and act accordingly. To understand their problems, the only way is through audio and visual. In deep forests, due to many trees we cannot monitor a large area. Whereas if we use a tool which recognizes syllables of various species, this tool can cover a large area with minimal equipment. This will save a ton of money for the government as well as rescue animals or birds when they are in emergency situation.

As our approach includes audio datasets with and without background noise, we can even take background noise into consideration and execute various methods in saving the environment as well. We can even record various background noise of forest fires, cutting down of trees with equipment. We can even take these into consideration and save environment.

In contrast to the research fields of Automatic Speech Recognition (ASR) and Music Information Retrieval (MIR), for which there exists extensive bibliography, the research field of Automatic Bird Identification (ABI) has received special attention only in the past decades (Tivarekar, 2017). Bird and Animal audio detection is one of the most intensive researches given by Detection and Classification of Acoustic Scenes and Events (DCASE 2018), since birds and animals are more easily detectable through the audio modality rather than vision. Traditional method used to recognize animal and bird sounds are from the generated spectrograms and continuous monitoring of those spectrograms for recognition of the species is a tedious task. Moreover, human decision is always subjective. Hence there is a great need for development of automated sound analysis.

"Bird sounds are divided into a set of hierarchical structure. The basic unit of sound produced by a bird is called as syllable. A repetitive group of syllables constitute a phrase. At times syllables and phrases are undistinguishable. A sequence of phrases constitutes a call." (Bang, 2018) Songs are complex and longer in duration than calls. However, these Automatic Bird Identification (ABI) tasks present various difficulties from the initial stage itself, since the acoustic sensors used for recording bird sound are subject to background noise such as wind and rain.

In this report, we will find an approach which is used to identify and detect birds and animal sounds with the help of pre-recorded datasets. To recognize the audio of species, we need to first know how the convolutional networks work in Tensor Flow.

To know about the convolutional networks, we need to learn in detail about the traditional Markov Model which is used to detect human speech commands. We will then use the similar Markov Model along with animal datasets instead of human. We will then convert all the .mp3 files to .wav files. All these .wav files are modulated in such a way that every audio file is having a similar frequency of 16 000 Hz and bit rate of 16-bit. These are then trained by machines and are recognized with the help of three main factors such as rate, cross entropy and accuracy. The clearer the dataset is, the higher accuracy will be.

To get a clear audio file without any background noise, we need to have pre-recorded sounds of all the possible background noise and make sure we cancel these noises in the considered dataset and after many approaches, we got an accuracy of 87.6% in detecting species sound.

Let's have a look at how audio recognition in species is important, some of the points are mentioned below:

• Firstly, the audio frequencies of birds carry a lot of information which can be either about climate change or about their habitats. With the help of various frequency ranges, we can help the species located at that habitat and save them if there are any problems. One other reason for change in frequencies might be due to lack of food. By acknowledging this kind of technique, we can save a lot of endangered species.

• Secondly, Continuous monitoring of species is a tedious task. It requires a lot of labor along with a lot of advanced equipment which causes organizations to spend a lot of money, time and labor. Instead, we can use this audio recognition tool in place of telescope and other video monitoring devices by installing the audio recognition tool in their place. With pre-recorded audio datasets, we can know if the species are in danger or not. A lot of money as well as time will be saved.

• Finally, we can protect the forests from forest fires as we will have pre-recorded audio dataset of forest fire. If this audio is recognized, then we can send rescue team on spot. We can even protect forests from deforestation. If there is any other sound, then other variable will be triggered and can send signals to the receiver. This can help forests along with the species, providing better and safe habitat for them to co-exist in.

## II. LITERATURE REVIEW

In this section, instead of covering broadly how machine learning is applied nowadays (e.g., internet of things (Zeng, E-AUA: An Efficient Anonymous User Authentication Protocol for Mobile IoT. IEEE Internet of Things Journal., 2018) (Zheng, 2014), social networks (Wang, 2018), activity recognition (Bhandari, 2017) (Pan, 2018), recommendation (Fu)), we will detailed information about the Audio Recognition done with the help of Tensor Flow.

"According to Sophia, Tensor Flow is used to implement complex DNN structures without getting complex mathematical details, and availability of large datasets. The machine learning model used in this paper is CNN which consists of three hidden layers." (Thakare, 2017) This algorithm predicts the bird sounds and gives output as 1 if found the audio clip or 0 if it couldn't find any. Spectrogram features are learned from the audio signals to detect bird audio. This network is implemented in Keras.

"Hassanali identifies and differentiates between bird's call and bird's song. These are separated with the help of syllable which is denoted as smallest part in audio." (Virani, 2017) The author depicts that recognition based on class probabilities and recording level detection is more suitable than syllable level detection. The frequency calculating technique used here was Mel Frequency Cepstral Coefficients (MFCC) which shows accurate results for audio recognition.

Arti V. Bang proposed various pattern techniques which are applied for sound classification such as "Mel Frequency Cepstral Coefficients (MFCCs), Gaussian Mixture Modeling (GMM) and Hidden Markov Model (HMM)" (Bang, 2018). The recordings obtained here are in different formats with different sampling rates and were casually recorded in a noisy environment. Various techniques such as Mean Computation, Principal Component Analysis along with K- Nearest Neighbors are used while testing the machine. The overall accuracy of bird's audio recognition kept increasing as k-folds are increased.

Sven Koitka identified a new technique wherein the audio files are preprocessed before being trained. They are extracted according to frequencies and then are examined with the help of bandpass filtering technique. "The output datasets are further undergone to noise filtering and silent region removal technique where author obtained pure audio datasets which contained only animal sound and are trained with Transfer Learning algorithm. To undergo this algorithm, the

convolutional neural network is finely-tuned and are trained from the scratch." (Koitka, 2017)

Dorota Kamiska and Artur Gmerek presented fully automated algorithm (kaminska, 2012). SOM and k-NN classifiers, have been chosen and compared in their paper for sounds of few species downloaded from various web sources (Disjoint sets with 70% - for preparing, 30% - for testing). The order precision for various highlights demonstrated that spectral features are the best for Automatic Species Recognition task. Their best outcomes had mean Classification precision of 69.94% with k-NN classifier and 52.92% with SOM classifier.

Chang-Hsing Lee et al. used frequency information to extract the syllables exactly (Lee, 2006). Averaged MFCCs in a syllable were used to identify species from their sounds. Experiments concluded that AMFCC greatly outperforms HMM and ALPC in training and testing. The average classification accuracy was up to 96.8% and 98.1% for frog calls and cricket calls, respectively.

Panu Somervuo, Aki Harma, and Seppo Fagerlund segmented a recording into individual syllables using a time-domain algorithm and segmented each region using three models such as Sinusoidal Model, Mel-Cestrum Model, and Descriptive Parameters. Dynamic time warping (DTW) algorithm was used for comparing variable length sequences (Somervu, 2006). Gaussian mixtures were used for modelling probability density functions in pattern recognition. The average recognition accuracy for single syllable was only around 40% to 50% depending on the method. The recognition results improved significantly in song-based recognition.

Iosif Mporas et al. evaluated the appropriateness of bird species recognition task with the help of real-field audio recordings of seven bird species, which are common for the Hymettus Mountain in Attica, Greece (Mporas, 2012). Two temporal and sixteen spectral audio descriptors computed using the open SMILE acoustic parameterization tool were used. For high SNR the boosting algorithm outperformed all the rest classification algorithms, while for low SNR the bagging meta- classifier offered slightly better performance than the boosting algorithm with maximum classification accuracy of 92.89%.

III. RELATED WORK

In the subsequent sections, we are going to provide a critical analysis of related works in the field of audio recognition using machine learning. Let's have a look at them below:

"Deep Learning based Bird Audio Detection":

In this paper, the author proposes a work which helps in detecting bird's audio based on deep learning algorithm using Tensor Flow. "Tensor Flow is used to implement complex DNN structures without getting complex mathematical details, and availability of large datasets. The machine learning model used in this paper is CNN which consists of three hidden layers." This algorithm predicts the bird sounds and gives output as 1 if found the audio clip or 0 if it couldn't find any. Spectrogram features are learned from the audio signals to detect bird audio. This network is implemented in Keras. This is inspired by human biological neurons. The processing neural signals along with deep learning is used in recognizing species audio. The preprocessed data is then mapped to the training data. By undergoing various normalizations through preprocessing functions, CNN is developed. The network is trained with back-propagation through time using Adam optimizer.

"Species Recognition Using Audio Processing Algorithm":

In this paper, the author identifies and differentiates between bird's call and bird's song. These are separated with the help of syllable which is denoted as smallest part in audio. The author depicts that recognition based on class probabilities and recording level detection is more suitable than syllable level detection. The frequency calculating technique used here was Mel Frequency Cepstral Coefficients (MFCC) which shows accurate results for audio recognition.

"Recognition of Bird Species from their Sounds using Data Reduction Techniques":

This paper illustrates various pattern techniques which are applied for sound classification such as Mel Frequency Cepstral Coefficients (MFCCs), Gaussian Mixture Modeling (GMM) and Hidden Markov Model (HMM). The recordings obtained here are in different formats with different sampling rates and were casually recorded in a noisy environment. Various techniques such as Mean Computation, Principal Component Analysis along with K- Nearest Neighbors are used while testing the machine. The overall accuracy of bird's audio recognition kept increasing as k-folds are increased.

"Recognizing Bird Species in Audio Files Using Transfer Learning":

In this paper, the author proposes a new technique wherein the audio files are preprocessed before being trained. They are extracted according to frequencies and then are examined with the help of bandpass filtering technique. The output datasets are further undergone to noise filtering and silent region removal technique where author obtained pure audio datasets which contained only animal sound and are trained with Transfer Learning algorithm. To undergo this algorithm, the convolutional neural network is finely-tuned and are trained from the scratch.

## IV. KEY RESEARCH AREAS

### A. Extracting Frequency Domain Representation

Recognizing animals and birds sound is a tough task for a machine than compared to human brain which recognizes without any stress. To train, we used two different datasets namely animals and birds. Both datasets contain short sound clips of various species. There is only a single event present in each sound file thus preventing overlapping.

### B. Using Animal Datasets

In this dataset, we took in three different animals naming the dataset folder as Bark for dog, Miaow for cat and Oink for pig. We've considered three animals so that we can work faster on a small dataset. The average duration of each audio dataset file is 1.0 s.

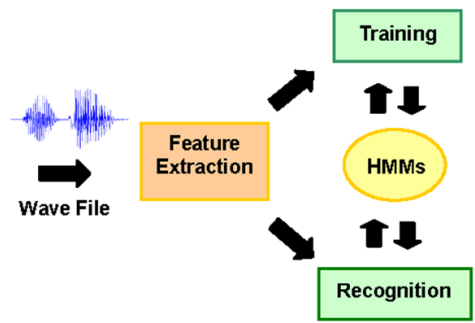

Fig. 1. Research Area Transition.

### C. Using Bird Datasets

In this dataset, we took in two different birds naming the dataset folder as Pigeon and Crow. The average duration for each audio dataset file is 1.0 s.

### D. Bandpass filtering

Most of the training datasets are under 500 Hz and above 18 000 Hz this causes a lot of problem while testing on different datasets and training outcomes will be inaccurate. To maintain the integrity among all the datasets, we need to increase and decrease the Hz of datasets and make every dataset equal to 16 000 Hz.

### E. Noise Filtering

Here we will reduce the background noise in the datasets by assuming few background noises. We take waterfall, heavy wind and many other scenarios into consideration and make sure we will reduce these background noises if found in the audio datasets so that we can train the machine with only animal or bird sound.

### F. Silent Region Removal

Few audio sources are found where we can hear animals and birds sound in an echo. This echoing effect will increase or decrease the frequency and modulation of voice and it will cause a lot of trouble in detecting the species. We use this technique to reduce such sounds.

## V. RESEARCH QUESTIONS

Identifying species by their sounds is important for animals and birds research and natural monitoring applications, particularly in identifying and finding the species. Besides, the greater part of the species sound recognition has developed to be species-specific. Furthermore, most of the animal vocalizations have evolved to be species-specific. Therefore, the utilization of animal and bird vocalizations to automatically identify their respective species is a natural and adequate way to environment monitoring, biodiversity assessment. A few research obstacles when managing species vocalizations are noisy data and label validity. The consolidation of noise data is vital when managing species vocalizations since accuracy is the main point, the datasets should be free from this noise. Detecting state of species is a tough job since researchers can just figure regarding what the creature is trying to communicate acoustically.

Some of the major questions that are addressed in audio recognition:

- "If waveforms are processed, what techniques can be adopted for signal processing?"
- "If spectrograms are processed, what techniques can be adopted for time-frequency domain?"
- "What are the useful acoustic components for animal call recognition?"
- "How can the developed species recognition system be evaluated?"

## VI. RESEARCH METHODOLOGIES

In this project, we will try and train a machine to recognize various animal voices for the betterment of the animals in our ecosystem to provide an environment better suitable for them and us. It is an arduous task for us to deploy human assets everywhere for the sole purpose of monitoring animals even after which it is susceptible to obvious errors. So, it is better to develop a machine which can do it all for us. This research will help start the whole process spoken about with detailed understanding of how to train the machine from scratch with very few exceptions. This can be used now during the preliminary times of development which will only improve the machine to be more effective as it learns further.

*A. Methodology*

After analyzing and understanding certain papers pertaining to the topic of research, the following are the steps I resorted to solve the issue at hand.

Traditional Markov Model was the technique used by many of the researchers to train the machine to recognize various variety of voices. Understanding this technique was the pivotal point to propel the project in the right direction. This gave me insight into how data sets are utilized to recognize syllables of a particular kind of sound or voice. However not a lot was being done just to recognize different species of animals and birds which was the issue at hand. Further researching on how to utilize the resources available to apply and enumerate possible ways to solve the problem was a conundrum.

After further research on the solutions already available and deducing the methodologies and algorithms used is only when we are able to find a feasible solution. The stark difference between the available solution and this process is obvious.

The course of the solution is clearly mentioned below using a flow chart. Every step in the flowchart is also clearly described. The importance of establishing such a working machine is also emphasized at the end.

As highlighted before, audio recognition using machine is a tedious task. There are a lot of complex algorithms and hidden layers performing various operations. Audio recognition of species is important for ecologists and to the researchers to calculate climate.

So, we are using the traditional Hidden Markov Model to recognize the audio of species.

The major research will be conducted by using a recognition model and this is the following structure:

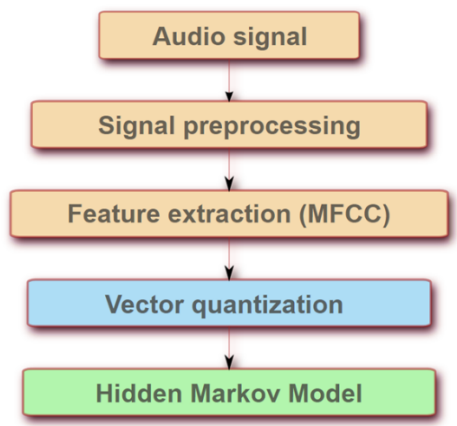

Fig. 2. Methodology Transition.

- Firstly, we will have few raw datasets. We need to convert all the raw datasets to .wav files and make sure they are ready for testing.
- Secondly, we will move to Signal Preprocessing stage where we will match all the frequencies and bandwidths of datasets. We need to change all datasets to the default bandwidth which is 16 000 Hz.
- Thirdly, we need to extract the data from the provided datasets and run with the provided sample background noise and make sure we will eliminate those from the dataset.
- Finally, we will Quantize and move on to Markov Model to decode the sound according to species. It will run the datasets with many normalizations and algorithms before generating the final confusion matrix.

VII. PROPOSED SOLUTION

In this part, we will discuss as to how we start categorizing animal data sets and recognizing the syllables produced by them. Below is a flowchart showing the whole process.

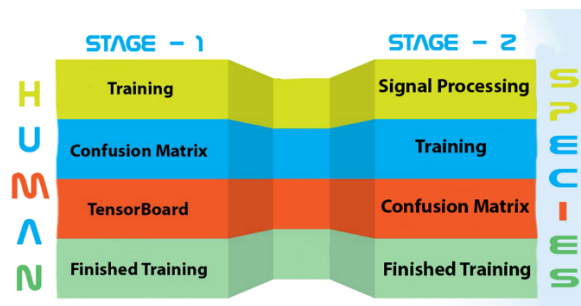

Fig. 3. Proposed Flowchart.

We begin by training the machine on how to recognize human voice by the help of pre-recorded data sets from Tensor Flow. Now we gradually start to add other data sets that were not in the pre-recorded data sets and test the performance of the trained machine in recognition. Once we succeed in that we start to add data sets of animal voices and train the machine to recognize them. During the training of animal/bird voice data sets we indulge in categorizing the different syllables produced by various animals/birds. The time quantum of these audio files is re-calibrated in a way that it becomes a more efficient methodology to detect the species of animals or birds involved in the training process.

### A. First Stage

Initially we found data sets containing human voices on Tensor Flow. The deep neural networks of Tensor Flow are considerably efficient as their processing speed is very high which leads to a very low processing time. We now understand the algorithm used on Tensor Flow to recognize and categorize human voice commands. The python file used to train data sets uses a traditional model called Markov model. Markov model considers three factors which are Rate, Accuracy, and Cross Entropy as parameters. It makes modulations to these parameters and checks at every 400 steps where each step is a different value assigned to Rate and Cross Entropy to form a confusion matrix. Every confusion matrix has different combination of Rate and Cross Entropy with which it is trying to determine the best way to achieve highest accuracy possible. After 18000 steps, it compares all the confusion matrices formed and deciphers the best combination of Rate and Cross Entropy where the Accuracy is at its highest.

Before executing following steps, we need to make sure that we've installed Python correctly and should make sure if the machine on which we are training can support GPU functionality or not. Generally, machines which support GPU functionality will have a capability to train faster than the machines which only support CPU functionality.

Step-1: We need to download the python files from the Tensor Flow website.
Step-2: Now we need to open command prompt and navigate to the program path and execute train code.
Step-3: As soon as we execute the train.py file, the Python program will download human command datasets from cloud repositories.

Step-4: As the download gets completed, we will now see the execution of the Python code where we will find series of trial and error method calculations.
For every 400 steps, out of 18 000 steps, we will get a confusion matrix as displayed below.

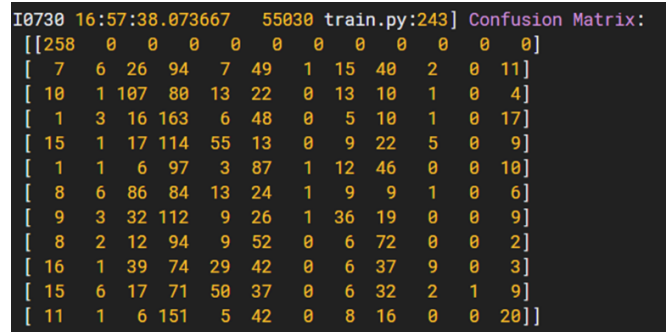

Fig. 4. Confusion Matrix.

This is known as Confusion Matrix. Each column represents a set of samples that were predicted to be each label, so first two rows belong to background noise and third row belongs to other sound. The remaining rows belong to individual speech command datasets ("yes", "forward", "backward" etc.).
Each row represents clips by their correct, ground truth labels. This matrix can be more useful than just a single accuracy score because it gives a good summary of what mistakes the network is making.
A perfect model would produce a confusion matrix where all the entries were zero apart from a diagonal line through the center. Spotting deviations from that pattern can help you figure out how the model is most easily confused, and once you've identified the problems you can address them by adding more data or cleaning up categories.
Step-5: As soon as 18 000 steps get executed, we will get final matrix along with the maximum possible detection accuracy of the provided datasets.
Step-6: We've now successfully trained machine and now we need to check if the machine is recognizing commands or not.

### VIII. METHOD AND ALGORITHM USED

HMM provides a simple and effective framework for modelling time-varying spectral vector sequences. Any system designed to work reliably as a real-world application must be robust because of the modulations in humans and another species of the ecosystem. HMM follow few stages to achieve the desired output.

1. Parameter Estimation: Different objective functions that are optimized in training and their effects on performance is estimated in this stage.
2. Adaptation and Normalization: This stage provides variety of techniques to achieve Robustness.
3. Noise Robustness: This stage handles convolutional noise and reduces background noise along with any echoes if there in audio datasets.

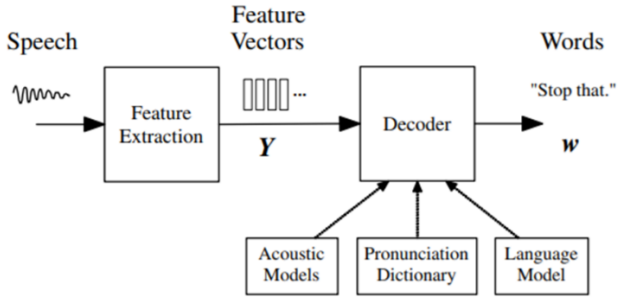

Fig. 5. Architecture of HMM

### A. Algorithm

The process begins in such a way by converting all the audio files from .mp3 to .wav. While the files are being converted, the audio waveform from microphone converts a series of fixed acoustic vectors $Y_{1:T} = y_1,\ldots,y_T$ and this process is known as feature extraction. Then there will be a decoder which tries to find the sequence of words $w_{1:L} = w_1,\ldots,w_L$ which will most likely generate Y.

$$\hat{w} = \arg\max_{w}\{p(Y|w)P(w)\}.$$

In this process, the acoustic model is not normalized. It is just analyzed as vectors and is decoded. It is often scaled by empirically determined constant. For any given sequence of words, this algorithm first converts all those series of words to phones. For example, the word "bat" is converted into /b/ /ae/ /t/. The parameters of these output phone are estimated for training dataset. The language model considered here is basically an N-gram model where each word will have N-1 predecessors.

### B. Feature Extraction

This stage provides actual representation of a waveform created by speech. This feature extraction method will minimize the loss of information by providing accurate distributional assumptions made by acoustic model. For all state output distributions, we used Gaussian distributions. This method is best suitable for convolutional networks.

All the encoding schemes are based on mel-frequency cepstral coefficients (MFCCs). These cepstral coefficients are considered to smooth the audio and helps in reducing the background noise. The smoothing of an FFT is done around 20 Frequency bins. The non-linear frequency scale used here is known as mel scale and this has a similar response to that of human ear.

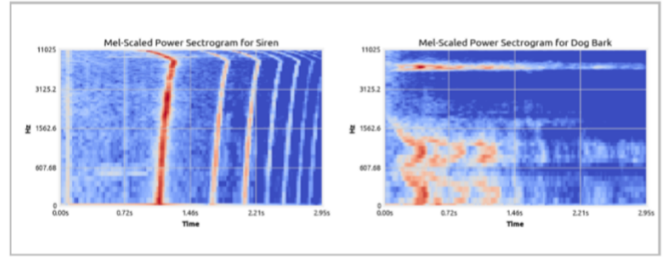

Fig. 6. Spectrogram Propagation.

Further psychoacoustic constraints are incorporated into a related encoding called perceptual linear prediction (PLP) (H, 1990). The linear prediction coefficients will compress power spectrum and will transform those coefficients to cepstral coefficients. In addition to these spectral coefficients, first order (delta) and second-order (delta–delta) regression coefficients attempt to compensate conditional independence which is an assumption made by HMM model. To detect the delta values, we need to undergo the following algorithm.

$$\Delta y_t^s = \frac{\sum_{i=1}^{n} w_i \left(y_{t+i}^s - y_{t-i}^s\right)}{2\sum_{i=1}^{n} w_i^2}$$

### C. Basic – Single Component

Every word is decomposed into sequence of small words known as base phones. Combination of these phones will lead to multiple pronunciation $p(Y|w) = \sum_{Q} p(Y|Q)P(Q|w)$, where summation of these valid pronunciations will produce a word. In practice, there will be various alternative pronunciations for every word summation.

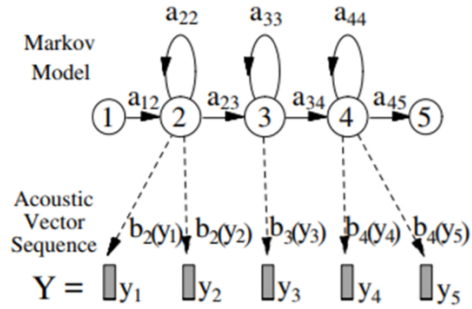

Fig. 7. Usage of Markov Model.

States are independent and keeps changing their values upon the given previous state. Observations will also variate according to the state change. Every state and observation will newly generate upon change of states. Composite HMM is formed by concatenating all the base phones and then acoustic livelihood is formed by the algorithm given below where θ is a state change sequence.

$$p(\boldsymbol{\theta},\boldsymbol{Y}|Q) = a_{\theta_0 \theta_1} \prod_{t=1}^{T} b_{\theta_t}(\boldsymbol{y}_t) a_{\theta_t \theta_{t+1}}.$$

This stage begins with base phones as initial parameters. After successive iterations of EM algorithm, the parameters improve the likelihood up to a maximum level. It then follows the Gaussian distribution to set all the transition probabilities to be the same. This gives a start to the model. The major problem with this model occurs while converting and decomposing nuances of the Vocabulistics of a language. The main problematic stage is where we use monophonic sounds. To resolve such problems, we need to add left and right neighbors of the word. We need to convert the monophonic sounds to tri-phonic sounds. The following approach will explain about the next stage.

All the states of leaf node are formed by cluster of previous parent nodes. The cluster of all states of nodes are formed as one pool node at the root node. All the leaf nodes are selected and are predetermined to maximize the likelihood of training data given. Gaussians at any node are calculated by the count of model parameters without referencing training data. We can even use greedy algorithm to increase the size of decision tree.

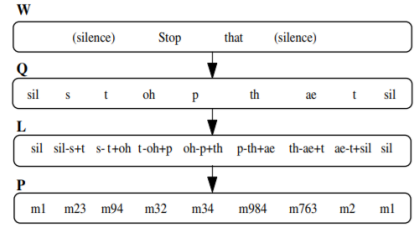

Fig. 2.3 Context dependent phone modelling.

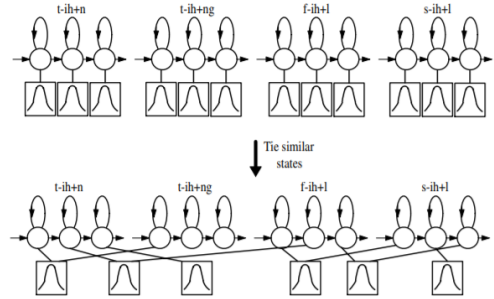

Fig. 2.4 Formation of tied-state phone models.

### B. Stage-2

Now, we know the process of training, testing and we also saw the algorithm behind audio detection. We need to download ideal datasets. Ideal datasets are those which are pre-recorded without any disturbance in spectral waves and there is no disturbance in background. The accuracy of the detection along with the displayed confusion matrix depends upon the datasets. The clearer the dataset, the higher is the accuracy we will achieve.

We need to make sure that all the audio files in the datasets are of similar frequency along with the bit rates. If one file is having 8-bit rate and if other file is having 16-bit rate, then we will have mel-frequency error where the files frequency will be mismatched. So, we need to make sure that all the files are

1. Set to 8-bit rate.
2. All the files are set as mono audio.
3. All the files should have a frequency range of 16 000 Hz.
4. The duration of each audio file should be 1 second.

The reason behind mono audio is that many audio files will have a similar bit rate and frequency. If their range is differentiating, then there will be a problem in creating spectrograms which are the main reason in detecting audio. If the files are having mono audio, then we need to leave as it is. If the files are having bi-audio, then we need to make sure that average value range is considered. To do this process, we need to convert all the .mp3 files to .wav files and during the converting process, we need to use some pre-defined tools which help in aligning all the datasets equally.

In our training program, we are considering our frequency range to be 16 000Hz. We need to make sure that we convert the files to that range or else there will be problem with mel-frequency variant.

We can even change the parameters if required. The reason behind considering audio files to be 1 second is due to the processing time and about the spectrograms. The accuracy of detection may vary depending upon change in time duration as well as the bit rate.

IX. TRAINING

A. Training-1

Let's consider altering the values from the training program and see the variation in accuracy. In this first training stage, we will consider the default parameters.

1. Resolution – 16 Bit
2. Sampling Rate – 16 000 Hz
3. Audio Channel – mono
4. Audio Trim – 1 Second

We are working on 6 datasets namely,

1. Silence
2. Unknown
3. Bark
4. Miaow
5. Pigeon
6. Peacock

Depending upon the dataset as well as the sampling rate, we can see that different confusion matrices will be formed as well as different spectrograms will be formed. After successful 18000 steps, we get validation accuracy to be 100%. And our final test accuracy to be 66.7%.

```
INFO:tensorflow:Step 18000: Validation accuracy = 100.0% (N=4)
INFO:tensorflow:Saving to "/tmp/speech_commands_train\conv.ckpt-18000"
INFO:tensorflow:set_size=9
INFO:tensorflow:Confusion Matrix:
 [[1 0 0 0 0 0]
  [0 0 0 0 0 0]
  [0 0 1 1 0 0]
  [0 0 0 1 0 1]
  [0 0 0 1 1 0]
  [0 0 0 0 0 2]]
INFO:tensorflow:Final test accuracy = 66.7% (N=9)
```

Fig. 8. Output of Training-1.

The confusion matrix is an ideal one when there is only 1's and 0's in it. As we can see, our confusion matrix is almost ideal and if we test on an audio then we will get its probability of recognition.

A.A Testing on different audio datasets

1. Peacock

```
D:\python\tensorflow-master\tensorflow\examples\speech_commands>python label
--labels=/tmp/speech_commands_train/conv_labels.txt \ --wav=/tmp/speech_data
2018-09-23 11:37:48.643489: I T:\src\github\tensorflow\tensorflow\core\plat
rts instructions that this TensorFlow binary was not compiled to use: AVX2
peacock (score = 1.00000)
miaow (score = 0.00000)
_silence_ (score = 0.00000)
```

Fig. 8.1. Testing on Peacock Sound.

As we can see, we got a perfect probability of 1 in detecting peacock audio. We can consider that this file is ideal for recognition of peacock species.

2. Bark

```
D:\python\tensorflow-master\tensorflow\examples\speech_commands>python label_wav.py \ --graph=/tmp/my_f
--labels=/tmp/speech_commands_train/conv_labels.txt \ --wav=/tmp/speech_dataset/bark/Dog_Sound_05.wav
2018-09-23 11:41:06.791298: I T:\src\github\tensorflow\tensorflow\core\platform\cpu_feature_guard.cc:14
rts instructions that this TensorFlow binary was not compiled to use: AVX2
bark (score = 0.99999)
_silence_ (score = 0.00000)
miaow (score = 0.00000)
```

Fig. 8.2. Testing on Dog Sound.

As we can see, we almost got a perfect probability of 0.9999 in detecting dog (bark) audio. We can consider that this file is ideal for recognition of dog species.

3. Miaow

```
D:\python\tensorflow-master\tensorflow\examples\speech_commands>python label_wav.py \ --graph=/tmp/my_f
--labels=/tmp/speech_commands_train/conv_labels.txt \ --wav=/tmp/speech_dataset/miaow/Cat_Sound_07.wav
2018-09-23 11:42:56.006926: I T:\src\github\tensorflow\tensorflow\core\platform\cpu_feature_guard.cc:14
rts instructions that this TensorFlow binary was not compiled to use: AVX2
miaow (score = 0.99606)
peacock (score = 0.00366)
pigeon (score = 0.00027)
```

Fig. 8.3. Testing on Cat Sound.

We can see, we almost got a perfect probability of 0.99 in detecting cat (meow) audio. We can consider that this file is ideal for recognition of cat species.

4. Pigeon

```
D:\python\tensorflow-master\tensorflow\examples\speech_commands>python label_wav.py \ --graph=/tmp/my_froze
--labels=/tmp/speech_commands_train/conv_labels.txt \ --wav=/tmp/speech_dataset/pigeon/Pigeon_Sound_02.wav
2018-09-23 11:44:37.215118: I T:\src\github\tensorflow\tensorflow\core\platform\cpu_feature_guard.cc:141] Y
rts instructions that this TensorFlow binary was not compiled to use: AVX2
pigeon (score = 0.99978)
bark (score = 0.00019)
miaow (score = 0.00003)
```

Fig. 8.4. Testing on Pigeon Sound.

We can see, we almost got a perfect probability of 0.99 in detecting pigeon audio. We can consider that this file is ideal for recognition of pigeon species.

By considering the probability of recognition, we can thus conclude that the audio datasets are ideal. The probability in detection is almost 1 so we need to now change the parameters and then see how the probability of detection as well as the final test accuracy rate differs.

*A.B. Spectrograms for tested audio datasets*

All the following spectrograms are generated with parameters as 16-bit rate, mono source, sampling rate as 16 000 Hz and are of duration 100 milliseconds.

1. Bark (Dog)

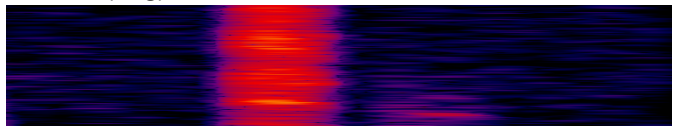

Fig. 8.1.1. Output Spectrogram on Dog Sound.

2. Miaow (Cat)

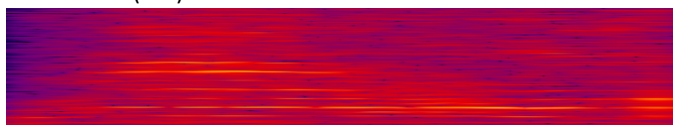

Fig. 8.1.2. Output Spectrogram on Cat Sound.

3. Pigeon

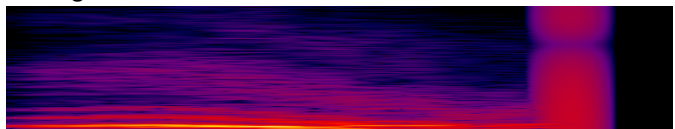

Fig. 8.1.3. Output Spectrogram on Pigeon Sound.

4. Peacock

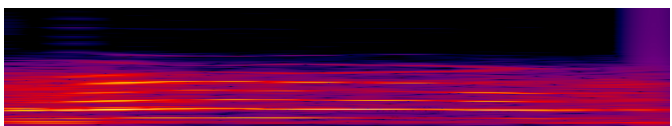

Fig. 8.1.4. Output Spectrogram on Peacock Sound.

*B. Training-2*

We are now changing the parameter values. We are altering our audio datasets and training our machine with following values.

1. Resolution – 16 Bit
2. Sampling Rate – 32 000 Hz
3. Audio Channel – mono
4. Audio Trim – 1 Second

```
INFO:tensorflow:Step 18000: Validation accuracy = 75.0% (N=4)
INFO:tensorflow:Saving to "/tmp/speech_commands_train\conv.ckpt-18000"
INFO:tensorflow:set_size=9
INFO:tensorflow:Confusion Matrix:
 [[1 0 0 0 0 0]
  [0 0 0 0 0 0]
  [0 0 1 1 0 0]
  [0 0 0 2 0 0]
  [0 0 0 1 1 0]
  [0 0 0 0 0 2]]
INFO:tensorflow:Final test accuracy = 77.8% (N=9)
```

Fig.9. Output Confusion Matrix of Test Case -2.

We are working on the previously used 6 datasets. As we've increased the frequency range which is the sampling rate to 32 000 Hz, we can see that the validation accuracy had massively dropped from 100% to 75%. We can also see that the test prediction accuracy has increased from 66.7% to 77.8% which is the positive case generated by this approach.

*B.A Testing on different audio datasets*

1. Peacock

```
D:\python\tensorflow-master\tensorflow\examples\speech_commands>python label_wav.py \ --graph=/tmp/my_frozen_g
--labels=/tmp/speech_commands_train/conv_labels.txt \ --wav=/tmp/speech_dataset/peacock/Peacock_Sound_01.wav
2018-09-23 20:54:53.330352: I T:\src\github\tensorflow\tensorflow\core\platform\cpu_feature_guard.cc:141] Your
rts instructions that this TensorFlow binary was not compiled to use: AVX2
miaow (score = 0.97768)
peacock (score = 0.01521)
_silence_ (score = 0.00710)
```

Fig. 9.1. Testing on Peacock Sound.

As we can see, we got a perfect probability of 1 in previous test case and we now have 0.97 as current probability to detect peacock audio. We can see a slight drop in the recognition of peacock species.

2. Bark

```
D:\python\tensorflow-master\tensorflow\examples\speech_commands>python label_wav.py \ --graph=/tmp/my_fro
--labels=/tmp/speech_commands_train/conv_labels.txt \ --wav=/tmp/speech_dataset/bark/Dog_Sound_05.wav
2018-09-23 20:55:41.732726: I T:\src\github\tensorflow\tensorflow\core\platform\cpu_feature_guard.cc:141]
rts instructions that this TensorFlow binary was not compiled to use: AVX2
miaow (score = 0.64611)
_silence_ (score = 0.22969)
bark (score = 0.11215)
```

Fig. 9.2. Testing on Dog Sound.

As we can see, we almost got a perfect probability of 0.9999 in previous test case and we now have a probability of 0.64 in detecting dog (bark) audio. We can see that there is a massive drop in recognition of dog species.

3. Miaow

```
D:\python\tensorflow-master\tensorflow\examples\speech_commands>python label_wav.py \ --graph=/tmp/my_frozen_gr
--labels=/tmp/speech_commands_train/conv_labels.txt \ --wav=/tmp/speech_dataset/miaow/Cat_Sound_07.wav
2018-09-23 20:56:19.968037: I T:\src\github\tensorflow\tensorflow\core\platform\cpu_feature_guard.cc:141] Your
rts instructions that this TensorFlow binary was not compiled to use: AVX2
miaow (score = 0.99768)
bark (score = 0.00178)
peacock (score = 0.00052)
```

Fig. 9.3. Testing on Cat Sound.

We can see, we almost got a perfect probability of 0.99 in previous as well as current test case in detecting cat (meow) audio. We can rule out this case in comparison of our test results.

4. Pigeon

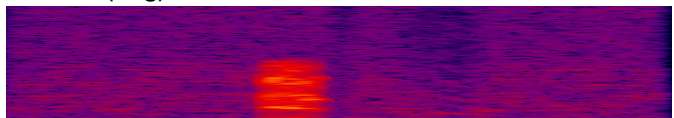

Fig. 9.4. Testing on Pigeon Sound.

We can see, we almost got a perfect probability of 0.99 in previous test case where as a massive drop in current test case can be observed resulting in probability of 0.81 to detect pigeon audio.

By comparing previous test case results and current test case results, we can conclude that increase in sample rate will increase the overall test accuracy but while testing on a current audio file, the recognition probability will massively decrease.

### B.B. Spectrograms for tested audio datasets

All the following spectrograms are generated with parameters as 16-bit rate, mono source, sampling rate as 32 000 Hz and are of duration 100 milliseconds.

1. Bark (Dog)

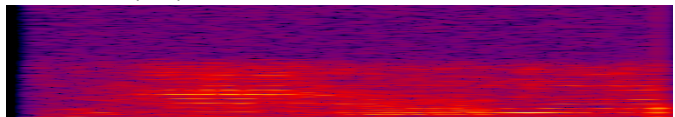

Fig. 9.1.1. Output Spectrogram of Dog Sound.

2. Miaow (Cat)

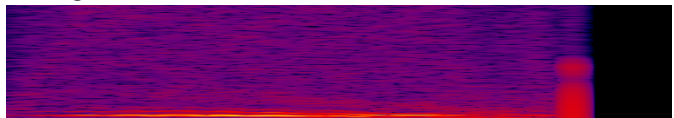

Fig. 9.1.2. Output Spectrogram of Cat Sound.

3. Pigeon

Fig. 9.1.3. Output Spectrogram of Pigeon Sound.

4. Peacock

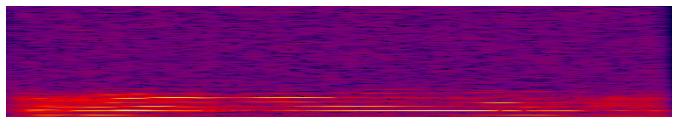

Fig. 9.1.4. Output Spectrogram of Peacock Sound.

### C. Training-3

We are now changing the parameter values. We are altering our audio datasets and training our machine with following values.

1. Resolution – 16 Bit
2. Sampling Rate – 32 000 Hz
3. Audio Channel – stereo
4. Audio Trim – 1 Second

Fig. 10. Confusion Matrix of Test Case-3.

We are working with the previously used 6 datasets. As we've increased the frequency range which is the sampling rate at 32 000 Hz and converted audio source from mono to stereo, we can see that the validation accuracy had maintained its state to be 75%. By this we can tell that changing the audio source from mono to stereo will not change the validation accuracy. We can also see that the test prediction accuracy has decreased from 77.8% to 66.7% which is considered as bad case generated by this approach.

### C.A Testing on different audio datasets

1. Peacock

Fig. 10.1. Testing on Peacock Sound.

As we can see, we got a perfect probability of 0.97 in previous test case and we now have 0.95 as current probability to detect peacock audio. We can see a slight drop in the recognition of peacock species.

2. Bark

Fig. 10.2. Testing on Dog Sound.

As we can see, we got a probability of 0.64 in previous test case and we now have a probability of 0.53 in detecting dog (bark) audio. We can see that there is a massive drop in recognition of dog species.

3. Miaow

Fig. 10.3. Testing on Cat Sound.

We can see, we almost got a perfect probability of 0.99 in previous as well as current test case in detecting cat (meow) audio. We can rule out this case in comparison of our test results.

4. Pigeon

Fig. 10.4.Testing on Pigeon Sound.

We can see, we got a probability of 0.81 in previous test case where there is a slight increase in current test case resulting in probability of 0.89 to detect pigeon audio.
By comparing previous test case results and current test case results, we can conclude that stereo audio source with increased sample rate will increase the overall test accuracy but while testing on a current audio file, the recognition probability will massively decrease.

*B.B. Spectrograms for tested audio datasets*

All the following spectrograms are generated with parameters as 16-bit rate, stereo source, sampling rate as 32 000 Hz and are of duration 100 milliseconds.

1. Bark (Dog)

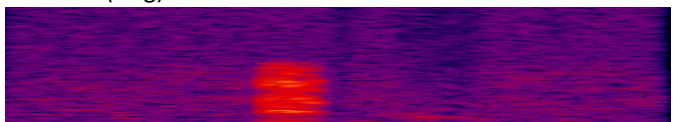

Fig. 10.1.1. Output Spectrogram of Dog Sound.

2. Miaow (Cat)

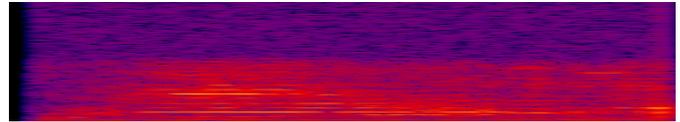

Fig. 10.1.2. Output Spectrogram of Cat Sound.

3. Pigeon

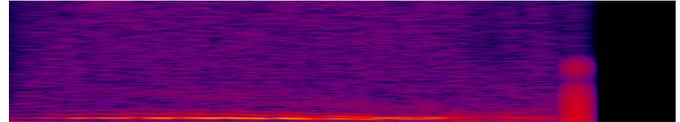

Fig. 10.1.3. Output Spectrogram of Pigeon Sound.

4. Peacock

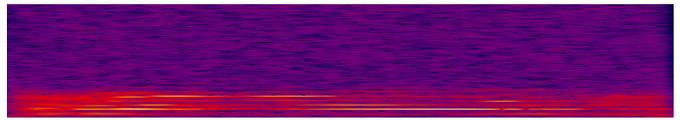

Fig. 10.1.4. Output Spectrogram of Peacock Sound.

*D. Training-4*

We are now changing the parameter values. We are altering our audio datasets and training our machine with following values.
1. Resolution – 16 Bit
2. Sampling Rate – 16 000 Hz
3. Audio Channel – stereo
4. Audio Trim – 1 Second

Fig. 11. Confusion Matrix of Test Case-4.

We are working on the previously used 6 datasets. As we've increased the frequency range which is the sampling rate to 16 000 Hz and we have also converted audio source from mono to stereo, we can see that the validation accuracy has its state changed from 75% to 100% which is a good start. By this we can tell that changing the audio source from mono to stereo along with change in sampling rate from 32 000 Hz to 16 000 Hz will affect the validation accuracy. We can also see that the test prediction accuracy has maintained its consistency at 66.7% which is generated by this approach.

*D.A Testing on different audio datasets*

1. Peacock

![Peacock test output]

Fig. 11.1. Testing on Peacock Sound.

As we can see, we got a probability of 0.95 in previous test case and we now have 0.99 as current probability to detect peacock audio. We can see a slight increase in the recognition of peacock species which is considered as a good start.

2. Bark

![Bark test output]

Fig. 11.2. Testing on Dog Sound.

As we can see, we got a probability of 0.64 in previous test case and we now have a probability of 0.66 in detecting dog (bark) audio. We can see that there is a slight increase in recognition of dog species.

3. Miaow

![Miaow test output]

Fig. 11.3. Testing on Cat Sound.

We can see, we almost got a perfect probability of 0.99 in previous case and there is a slight decrease which results in 0.93 as probability in recognition of cat (meow) audio in this case. This is considered as bad case scenario in the field of recognition.

4. Pigeon

![Pigeon test output]

Fig. 11.4. Testing on Pigeon Sound.

We can see, we got a probability of 0.89 in previous test case where there is a slight increase in current test case resulting in probability of 0.99 to detect pigeon audio.

By comparing previous test case results and current test case results, we can conclude that stereo audio source with decrease in sample rate will increase the overall test accuracy but while testing on a current audio file, the recognition probability will massively decrease.

*D.B. Spectrograms for tested audio datasets*

All the following spectrograms are generated with parameters as 16-bit rate, stereo source, sampling rate as 16 000 Hz and are of duration 100 milliseconds.

1. Bark (Dog)

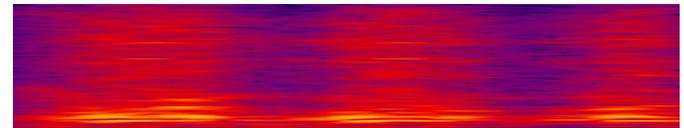

Fig. 11.1.1. Output Spectrogram of Dog Sound.

2. Miaow (Cat)

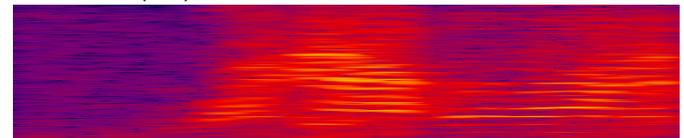

Fig. 11.1.2. Output Spectrogram of Cat Sound.

3. Pigeon

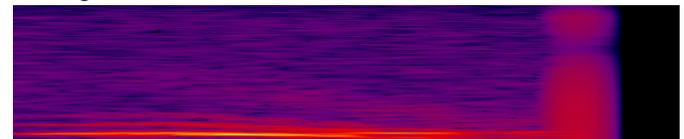

Fig. 11.1.3. Output Spectrogram of Pigeon Sound.

4. Peacock

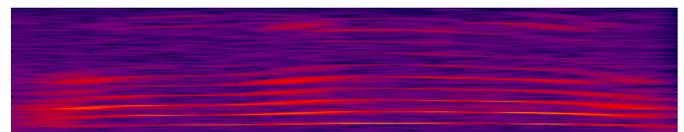

Fig. 11.1.4. Output Spectrogram of Peacock sound.

## X. FRAMEWORK

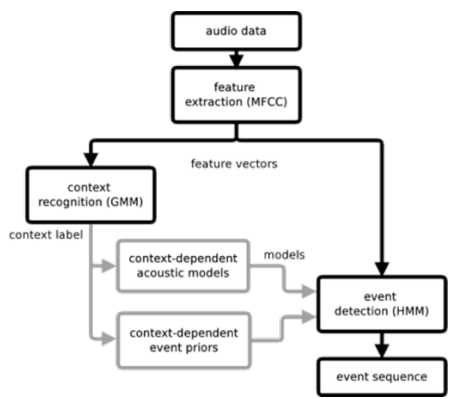

Fig. 12. Application Architecture.

Given above is the application architecture of audio recognition in the field of machine learning. It consists of five major components which are "audio data", "Feature Extraction", "Context Recognition", "Event Detection (HMM)" and "Event Sequence".

The audio data is considered as the pre-recorded input datasets which are used for testing. These audio datasets play an important role in recognition of species. All the validation accuracy along with the target accuracy is estimated depending upon the audio we choose. The algorithm might give various outputs depending upon the input which is the "audio data".

These audio datasets are then moved to "feature extraction" process. This process will represent the waveform created by the audio dataset. This method will minimize the loss of information and provides accurate assumptions depending upon the acoustic model we use. This method is best suitable for convolutional networks.

Before moving on to the next stage which is either context recognition or even detection, the feature extraction process will run through many different stages such as "pitch modelling" or "prediction".

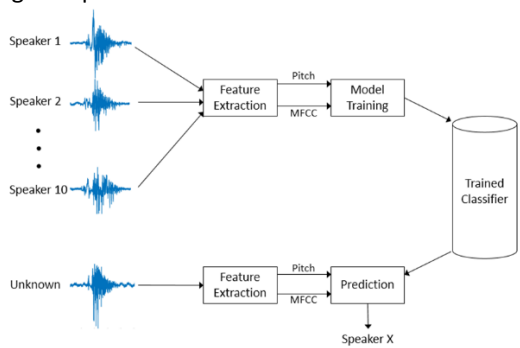

Fig. 13. Interior Working of Audio Recognition.

Model Training is the process which is generally executed if the dataset is good and is free of background noise. All the accurately recognized sounds will be moved on to the model training stage wherein this model training stage will train the machine to recognize such voices without any trouble. These audio datasets can either be human voice or sounds of species. Until and unless the datasets are good for training purpose, the model training will have no problem in training the machine.

As soon as the training is done, it is then moved to the trained classifier dataset. This will record all the occurrences and the accuracy probability to detect an audio. This is usually shown to the users in the form of confusion matrix.

If there is any small trouble with the audio, which might be the case where the background noise is more than the audio or there are other sounds which are disturbing the audio, along with the trained classifier audio datasets are sent to prediction stage. This prediction stage will take these values and make sure that it recognizes the background noise and then removes it.

As soon as training is done, the output of those files is then sent to event detection stage where Hidden Markov Model (HMM) method will be applied. This HMM will make sure to reduce the error and reduces the background noise. This algorithm is used to detect the audio after the training process and allocate a tag for them.

All the final outputs are then recorded in the form of ". ckpt" file. These files will store all the event sequences which are generated after the training process. These files are generated regularly after every 400 steps. These event sequences store all the information about the validation accuracy along with the total accuracy. This will also record the confusion matrix generated at that stage. To view output of final matrix along with the detection, we can just access this file and look at the Tensor Board.

## XI. DISCUSSION OF RESULTS

In order to prove our method, we have chosen 4 different species with 20 audio datasets each. All of these audio datasets are pre-recorded wherein few of these audio datasets have background noise and few don't. The overall accuracy is decided by three parameters such as frequency, audio rate, sampling channel. All of these are recorded with a time length of 100 milliseconds.

We will now test which method along with which parameters are best suitable to train data. We changed the values of audio rate, frequency along with the sampling channel. We will see the change in validation accuracy along with change in overall accuracy.

| Test Cases | Sampling Channel | Frequency | Rate |
|---|---|---|---|
| Test Case1 | Mono | 16 000 Hz | 16 |
| Test Case2 | Stereo | 16 000 Hz | 16 |
| Test Case3 | Mono | 32 000 Hz | 16 |
| Test Case4 | Stereo | 32 000 Hz | 16 |

Table. 1. Considered Parameters.

| Validation Accuracy | Overall Accuracy |
|---|---|
| 100 | 66.7 |
| 100 | 66.7 |
| 75 | 77.8 |
| 75 | 66.7 |

Table. 2. Output Results.

As we can see above, the test case-1 and test case-2 are having perfect validation accuracy with good overall accuracy of 66.7%. We need to now compare the test cases while testing on a particular audio data file. The prediction accuracy is the only way in which we can set the ideal parameters which can be used to train the machine to recognize audio of species.

| Test Cases | Dog | Cat | Pigeon | Peacock |
|---|---|---|---|---|
| Test Case1 | 0.99 | 0.99 | 0.99 | 1 |
| Test Case2 | 0.11 | 0.99 | -- | 0.015 |
| Test Case3 | 0.12 | 0.99 | -- | 0.009 |
| Test Case4 | 0.66 | 0.01 | 0.99 | 0.99 |

Table. 3. Testing on Individual Audio Datasets.

By considering the above displayed results generated by test cases, we can clearly deduce that testcase-1 is the best approach to recognize the audio datasets and by considering this approach we get the total accuracy as 66.7%.

So therefore, we can observe that our method is much more effective when compared to other audio recognition test cases that are considered.

XII. PROBLEM ANALYSIS

Although previously we have proved that our method is the best approach used in recognizing the audio of species, given below are few assumptions that we have considered for our method:

This recognition ability is tested on few audio datasets which aren't same in any syllable. All the accuracies have only been recorded for limited set of species. Let us assume two species which are fox and dogs. Both species sound the same while howling. There will be the slightest difference in some ranging syllable which cannot be detected by humans. This method might be able to generate spectrograms of both species howling. But it might also get confused at some point. To reduce this kind of problems, we can either create an audio recognition for species which are domestic separately, or we can even increase the time duration of both audios and then find difference while lowering their voices.

Secondly, if we consider Parrot which is a bird species, sounds like humans and another objects voice. This will confuse the machine to recognize if it is human's voice or the audio of parrot. It can even reciprocate background noise of various audio datasets wherein we will get the output recognized audio as "other" or "unknown". To reduce this type of problems, we can have an approach where we can see the pitch variation difference in each syllable and find out if it is parrot or human or any other destruction caused to the forest.

Finally, all the audio datasets in this method are predefined to detect them. If there is any case where we can update the audio datasets with lot of possibility audio produced by respective species, then this will help in recognizing the species without any loss in validation accuracy as well as in overall accuracy. So therefore, we assume that increase in various probabilistic sounds in audio datasets along with increase in time duration of every individual audio of species will help in detecting species perfectly without any loss.

## XIII. Conclusion

In this report, we have given a brief overview about the audio recognition of species and how they are helpful in real lives. We also provided the usage and working of commonly used traditional Hidden Markov Model. We have provided different stages of execution along with the detail approach which helped in recognizing audio. We've chosen the best approach which is used to recognize the audio and our method still needs improvement such as decreasing or increasing the bit rate along with the increase in audio duration. This can help in improving the overall accuracy even though this method is successfully recognizing the audio without any trouble. So, in the end our model can recognize the audio of species without any errors. Our work can serve as a foundation for others to continue our line of work and resolve interesting research challenges saving species from being endangered.

## XIV. Future Directions

Even with the extensive research on the approaches to recognize animal voice, there are many ways in which machines fail to recognize them accurately. Every proposed model so far just had an accuracy of about 66%. The challenges faced by audio recognition is more and by developing an accurate algorithm, we can detect species and help them from being endangered. We can even add various background sounds such as forest fires, breaking down of trees which helps to easily recognize the forest fires before it causes serious damage. Audio is the main weapon which can cover a large area than that of video where the vision of the camera will be interrupted by the trees. Further recognition techniques can also be added into this architecture to make it even more effective than others.